\documentclass[sigconf,screen, pbalance]{acmart}

\usepackage{booktabs}
\usepackage{multirow}
\usepackage{siunitx}

\usepackage[utf8]{inputenc}
\usepackage{xcolor}
\usepackage{xspace}
\usepackage{minted}
\usepackage{wrapfig}
\usepackage{algpseudocode}
\usepackage{algorithm}  
\usepackage{algorithmicx}
\usepackage{caption}
\usepackage{mathtools}
\usepackage{etc}
\usepackage{adjustbox}
\usepackage{tabularx}

\usepackage{tcolorbox}
\usepackage{listings}
\usepackage{url}
\usepackage[dvipsnames]{xcolor}
\usepackage{subcaption}
\setcopyright{none}
\newcommand{\se}{\textsc{se}}

\AtBeginDocument{%
  \providecommand\BibTeX{{%
    \normalfont B\kern-0.5em{\scshape i\kern-0.25em b}\kern-0.8em\TeX}}}

\usepackage{enumitem}
\setlist{topsep=0pt, leftmargin=*}
\setcopyright{none}
\acmBooktitle{Companion Proceedings of the 34th ACM Symposium on the Foundations of Software Engineering (FSE '26), June 5--9, 2026, Montreal, Canada}
\acmConference[FSE '26]{34th ACM Symposium on the Foundations of Software Engineering}{June 5--9, 2026}{Montreal, Canada}

\renewcommand\footnotetextcopyrightpermission[1]{} % removes footnote with conference information in first column
\title{Energy Flow Graph: Modeling Software Energy Consumption}
\author{Saurabhsingh Rajput}
\affiliation{
  \institution{Dalhousie University}
  \city{Halifax}
  \country{Canada}}
\email{saurabh@dal.ca}

\author{Tushar Sharma}
\affiliation{
  \institution{Dalhousie University}
  \city{Halifax}
  \country{Canada}}
\email{tushar@dal.ca}
\begin{document}

\begin{abstract}
The growing energy demands of computational systems necessitate a fundamental shift from performance-centric design to one that treats energy consumption as one of the primary design considerations. Current approaches treat energy consumption as an aggregate, deterministic property, overlooking the path-dependent nature of computation, where different execution paths through the same software consume dramatically different energy. We introduce the \textbf{Energy Flow Graph (\efg)}, a formal model that represents computational processes as state--transition systems with energy costs for both states and transitions. \efg enables various applications in software engineering, including static analysis of energy-optimal execution paths and a multiplicative cascade model that predicts combined optimization effects without exhaustive testing. 
Our early experiments demonstrate \efg's versatility across domains: in software programs validated through 3.5 million executions, 15.6\% of solutions exhibit high path-dependent variance (CV $>$ 0.1), while structural optimization reveals up to 705$\times$ energy reduction. In AI pipelines, the cascade model predicts optimization combinations within 5.1\% error, enabling selection from 4.2 million possibilities using only 22 measurements.
The \efg transforms energy optimization from trial-and-error to systematic analysis, providing a foundation for green software engineering across computational domains.
\end{abstract}


\maketitle

% \vspace{-2mm}
\section{Introduction}
The growing energy demands of computation---from mobile devices to massive AI training clusters---have made energy efficiency a primary design constraint~\cite{verdecchia2023systematic, georgiou2022green, rajput2024enhancing}. The emerging fields of \textit{green AI} and \textit{green software engineering} aim to optimize software energy consumption~\cite{schwartz2020green, manotas2016empirical, lee2024survey}. However, current approaches---whether high-level profiling, hardware performance counters, or micro-benchmarking---share a critical limitation: they treat energy as an aggregate, deterministic attribute of the overall process. This perspective overlooks the intrinsic path-dependent nature of software execution, in which the actual energy consumed varies significantly with the specific sequence of state transitions determined by inputs, runtime conditions, and hardware states.

A fundamental disconnect exists in current methodology: while execution is dynamic and path-dependent, standard energy models remain static and aggregate. Current practices, such as annotating code segments with average measurements~\cite{chung2025mlenergy,rajput2024enhancing}, fail to distinguish between execution paths, mask state-transition costs, and provide no insight into how optimizations interact. 
While some approaches employ state machines for energy analysis~\cite{duarte2019model,baier2014probabilistic} or trace-based profiling~\cite{gomes2022model}, they lack either the path-dependent optimization mechanisms or the thermodynamic foundations necessary for principled energy reasoning. 
For instance, database queries executing identical plans may consume radically different energy depending on join orders or data locality---nuances lost in aggregate annotations. Similarly, the energy footprints of machine learning (ML) models vary dramatically depending on input characteristics and batch sizes. 
Without path-level visibility, developers cannot identify energy-intensive execution patterns or predict how code changes affect energy behavior across workloads.
This path-insensitive modeling creates a distinct but related challenge: the combinatorial explosion of optimization interactions.
When $n$ optimizations are available, practitioners must choose from $2^n$ possible combinations. Measuring individual optimizations A and B at 20\% and 30\% savings provides no basis to predict their combined effect---whether multiplicative, additive, or antagonistic~\cite{rajput2025tu}---forcing either exhaustive testing, deployment of suboptimal single-optimization strategies or abandoned optimization efforts.

These practical challenges---\textit{path insensitivity, hidden transition costs, and combinatorial uncertainty}---stem from a deeper gap: the lack of a formal model connecting abstract software states--transitions to thermodynamic costs. While Landauer's principle~\cite{landauer1961irreversibility} establishes that logical state transitions fundamentally drive physical energy dissipation, current software engineering (\se{}) treats energy as an abstract metric rather than a physical consequence of computation. Bridging this gap between logical flow and thermodynamic reality is essential for developing robust optimization frameworks.

To address these challenges, we introduce the \textbf{Energy Flow Graph} (\efg), a computational model that bridges \emph{graph theory}, \emph{software engineering}, and \emph{thermodynamic} principles. The \efg abstracts any finite computational process as a state--transition system with explicit energy annotations for both states and transitions. This unified representation enables energy analysis across the system lifecycle: \textit{before} execution for design-time optimization planning, \textit{during} runtime for adaptive decision-making, and \textit{after} execution for post-mortem diagnosis of energy behavior. By modeling both potential paths (pre-execution) and activated paths (post-execution), \efg supports development of optimization strategies grounded in the underlying computational structure, including static analysis to identify energy-optimal paths, predictive models to forecast the combined effects of optimization strategies, and runtime adaptation based on path activation patterns. 

In summary, this work makes the following \textbf{contributions}.
First,
we introduce and define \efg{}, creating a bridge between graph theory, software engineering, and thermodynamic principles. 
Second, we demonstrate \efg's utility and versatility as an analytical tool for energy optimizations using two large-scale case studies. The replication package is available online~\cite{4openAnonymousGithub}.

% \vspace{-2mm}

\section{Energy Flow Graph (\efg)}
\label{sec:efg}

\setlength{\abovedisplayskip}{1pt}
\setlength{\belowdisplayskip}{1pt}
\setlength{\abovedisplayshortskip}{1pt}
\setlength{\belowdisplayshortskip}{1pt}

Energy Flow Graph (\efg) is a formal, domain-agnostic model that represents computational software processes
as state--transition systems annotated with energy costs, enabling systematic reasoning about and optimizing their path-dependent energy consumption.

% \vspace{-4mm}
\subsection{Definition}

\begin{definition}[Energy Flow Graph]
\label{def:efg}

An Energy Flow Graph is a directed graph formalized as a tuple $\mathcal{G} = (V, E, \mathcal{C}_s, \mathcal{C}_t, \mathcal{P})$, where:
\begin{itemize}
    \item $V$ is finite set of \textbf{vertices} representing discrete computational \textbf{states}.
    \item $E \subseteq V \times V$ is a set of directed \textbf{edges} representing \textbf{transitions} between states.
    \item $\mathcal{C}_s: V \to \mathbb{R}_{\geq 0}$ is the \textbf{state cost function}, assigning energy cost to execute each state.
    \item $\mathcal{C}_t: E \to \mathbb{R}_{\geq 0}$ is the \textbf{transition cost function}, assigning energy overhead to each state transition.
    \item $\mathcal{P}: E \to [0,1]$ is the \textbf{transition probability function} (optional, for stochastic analysis), where for each vertex $v \in V$, $\sum_{(v,v') \in E} \mathcal{P}(v,v') \leq 1$.
\end{itemize}

A \textbf{valid execution path} $\pi$ from vertex $v_1$ to vertex $v_k$ is a sequence of vertices $\pi = (v_1, v_2, \ldots, v_k)$ such that $(v_i, v_{i+1}) \in E$ for all $i \in \{1, \ldots, k-1\}$.

The \textbf{total energy} of path $\pi$ is:
\begin{equation}
E_{\text{total}}(\pi) = \sum_{i=1}^{k} \mathcal{C}_s(v_i) + \sum_{i=1}^{k-1} \mathcal{C}_t(v_i, v_{i+1})
\label{eq:path_energy}
\end{equation}
% \vspace{-2mm}
\end{definition}

This formulation explicitly captures the path-dependent nature of energy consumption: different execution paths through the same code consume different energy based on the states and transitions activated. \textbf{Due to space constraints, detailed derivations for equations are provided in the replication package~\cite{4openAnonymousGithubDerivation}}.

% \vspace{-2mm}
\subsection{Energy Analysis Methods}
\label{subsec:analysis_methods}

The \efg framework supports two distinct modes of analysis depending on whether the execution path is fully deterministic or governed by probabilistic uncertainty.

\subsubsection{Deterministic Path Analysis (Bounds \& Traces)}
\label{eq:bounds}
This mode applies when analyzing a specific execution trace (\eg{} post-mortem profiling) or establishing safety bounds. Since energy costs $\mathcal{C}_s$ and $\mathcal{C}_t$ are non-negative, identifying energy limits is reducible to a shortest/longest path problem on a weighted graph. The Best-Case Energy Consumption (BCEC) and Worst-Case Energy Consumption (WCEC) are naturally defined as $\text{BCEC} = \min_{\pi \in \Pi} E_{\text{total}}(\pi)$ and $\text{WCEC} = \max_{\pi \in \Pi} E_{\text{total}}(\pi)$, where $\Pi$ is the set of all valid paths from entry to terminal states.
For acyclic graphs, these bounds are computable using standard shortest-path algorithms~\cite{cormen2022introduction}.
These bounds are valuable for energy budgeting in resource-constrained, real-time systems.

\subsubsection{Stochastic Path Analysis}
When the precise execution path is not known a priori (\eg{} input-dependent control flow), we model the system as a Markov Chain~\cite{baier2014probabilistic}, where future states depend only on the current state and transition probabilities $\mathcal{P}$, not on execution history. We distinguish between \textit{evaluating} a fixed system and \textit{optimizing} a controllable one.

\noindent \textbf{Expected Energy (Fixed Policy):} For standard software analysis where transition probabilities are fixed by input distributions(\eg{} profiled from representative workloads), the goal is to compute the \textit{Expected Energy Consumption} $E_{\text{exp}}(v)$ from state $v$. 
This is given by the Bellman equation~\cite{bellman1952theory} for a Markov Decision Process (MDP)~\cite{bellman1966dynamic}:
\begin{equation}
E_{\text{exp}}(v) = \mathcal{C}_s(v) + \sum_{(v,v_j) \in E} \mathcal{P}(v,v_j) \left[\mathcal{C}_t(v,v_j) + E_{\text{exp}}(v_j)\right]
\label{eq:expected_energy}
\end{equation}
with the boundary condition $E_{\text{exp}}(t) = \mathcal{C}_s(t)$ for terminal states. For finite state spaces, this yields a linear solvable system~\cite{littman2013complexity}.

\noindent \textbf{Policy Optimization:} When the system can choose transitions at runtime (\eg{} adaptive algorithm selection, approximate computing, compiler optimization choices), the objective becomes finding an energy-minimizing policy. Instead of probability-weighted sums (as Eq~\ref{eq:expected_energy}), we select the minimum-cost successor at each state, yielding the MDP formulation:

\begin{equation}
E_{\text{opt}}(v) = \mathcal{C}_s(v) + \min_{(v,v_j) \in E} \left[\mathcal{C}_t(v,v_j) + E_{\text{opt}}(v_j)\right]
\label{eq:policy_opt}
\end{equation}

The optimal policy $\pi^*(v) = \arg\min_{(v,v_j) \in E} [\mathcal{C}_t(v,v_j) + E_{\text{opt}}(v_j)]$ identifies the energy-minimizing transition at each state, computable via dynamic programming~\cite{bellman1966dynamic}.

% \vspace{-2mm}

% \vspace{-4mm}
\subsection{Energy Optimization Constructs}
\label{subsec:optimization_constructs}

\efg enables systematic energy optimization through theoretical constructs grounded in graph theory and probability~\cite{bellman1952theory, bellman1966dynamic}.

\noindent \textbf{Node Optimization Impact:} Local optimizations to individual nodes create cascading effects throughout the graph. The global impact of optimizing node $v_i$ from energy cost $E_i$ to $E'_i$ is the probability-weighted sum of per-path savings multiplied by execution frequency~\cite{tiwari2002power}:
\begin{equation}
\Delta E_{\text{global}} = \sum_{\pi \in \Pi(v_i)} \Pr(\pi) \cdot (E_i - E'_i) \cdot \text{freq}(v_i, \pi)
\label{eq:node_impact}
\end{equation}

where $\Pi(v_i)$ is the set of paths containing $v_i$, $\Pr(\pi)$ is the execution probability of path $\pi$, and $\text{freq}(v_i, \pi)$ is the execution frequency of $v_i$ within path $\pi$. Parameters are derived from representative profiling data or symbolic execution bounds. 
This formulation reveals that high-frequency nodes in frequently executed paths provide disproportionate optimization leverage. 

\noindent \textbf{Multiplicative Cascade Model:} For orthogonal optimizations targeting independent computational resources (\eg{} one reduces CPU cycles, another reduces memory bandwidth), energy savings compound multiplicatively rather than additively~\cite{4openAnonymousGithubDerivation}:
\begin{equation}
S_{\text{combined}} = 1 - \prod_{i=1}^{n} (1 - s_i)
\label{eq:multiplicative_model}
\end{equation}
where $s_i$ represents the fractional energy saving of optimization $i$. This model enables predicting combined optimization effects without exhaustive testing, addressing the combinatorial explosion problem in optimization space exploration.

\noindent \textbf{Hierarchical Composition:} \efg supports multi-scale analysis through compositional abstraction. A macro-state $v_{\text{macro}}$ can represent an entire sub-graph $\mathcal{G}_{\text{micro}} = (V_{\text{micro}}, E_{\text{micro}})$, with the relationship:
\begin{equation}
\mathcal{C}_s(v_{\text{macro}}) = \sum_{v \in V_{\text{micro}}} \mathcal{C}_s(v) + \sum_{e \in E_{\text{micro}}} \mathcal{C}_t(e)
\label{eq:hierarchical}
\end{equation}

This compositional property ensures energy conservation across abstraction scales and enables modular analysis where optimizations at one level propagate predictably through the hierarchy.

% \vspace{-2mm}

\section{Applications}
\label{sec:validation}

To demonstrate \efg's utility as a predictive tool across computational scales, we present two case studies that validate its core constructs through empirical analysis. These applications map directly to the formal capabilities defined in Section~\ref{sec:efg}.

\noindent \textbf{Case Study 1} applies \efg to algorithmic software, demonstrating path-dependent energy variance and structural optimization potential through $\approx 3.5$ million cycle-accurate simulation of $39744$ algorithmic kernels. \textbf{Case Study 2} applies \efg to AI pipelines, validating the multiplicative cascade model for predicting optimization combinations of AI pipelines.

\subsection{Case Study 1: Algorithmic Energy Analysis}
\label{subsec:rq1}

\textbf{Application Context.} Developers face a fundamental challenge: predicting which algorithmic implementations and input patterns will minimize energy consumption. \efg addresses this by modeling algorithms as state-transition systems where energy varies by execution path. We validate whether \efg's constructs~\ref{sec:efg}---transition probabilities $\mathcal{P}$, state costs $\mathcal{C}_s$, and structural topology---accurately capture and explain observed energy behaviors.

\noindent \textbf{Dataset.} We use programs from Shypula \etal{}~\cite{shypula2023learning} with $3,507,435$ verified executions across $1474$ problems and $39,744$ unique solutions, each evaluated on an average of $88.3$ test inputs. Records capture energy (Joules), instructions, cycles, power (Watts), and IPC. For each solution, we conceptually construct a statement-level \efg where vertices represent states and edges represent control flow; different inputs activate distinct paths $\pi$, leading to measurable energy differences $E_{\text{total}}(\pi)$ (Eq.~\ref{eq:path_energy}).

\noindent \textbf{Measurement.} We used Sniper cycle-accurate simulator~\cite{heirman2012sniper} with McPAT power model~\cite{li2009mcpat}, modeling AMD EPYC 9554P (Zen 4, 5nm, 1.5 GHz). Simulation ensures: (1) determinism ($E(\pi)$ depends strictly on path $\pi$), (2) microarchitectural fidelity (captures transition costs $\mathcal{C}_t$), and (3) reproducibility.

\begin{figure}[t]

    \centering
    % \vspace{-4mm}
    \includegraphics[width=\columnwidth]{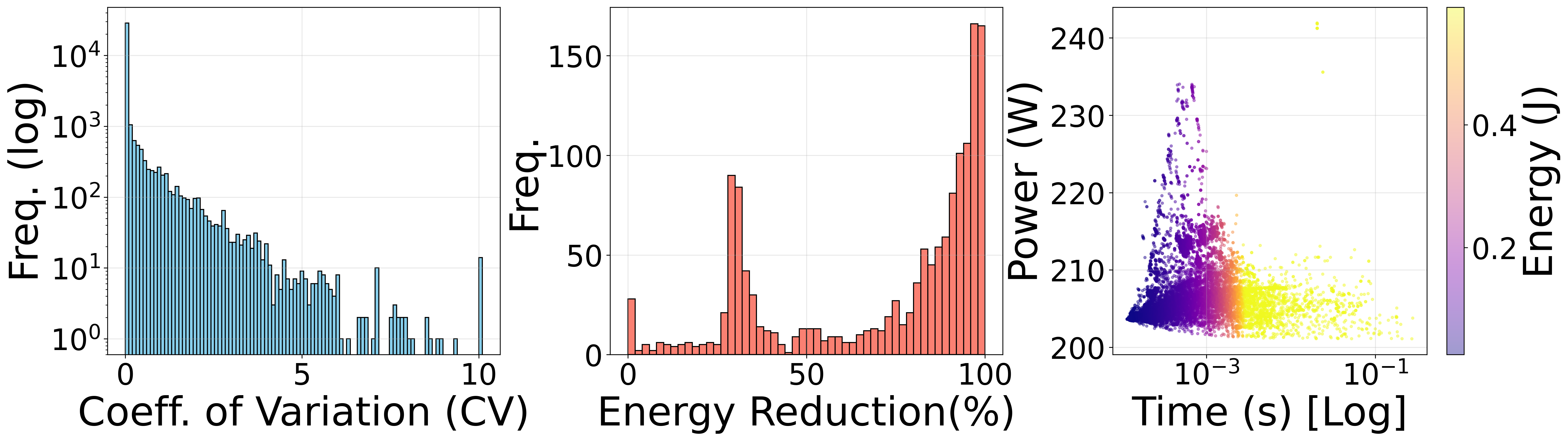}
    % \vspace{-3mm}
    \caption{Energy variance analysis. (a) Path dependency: intra-solution CV distribution; 15.6\% with $CV > 0.1$ validate stochastic modeling. (b) Structural impact: 42\% of problems offer $>90\%$ energy reduction via topology selection. (c) Thermodynamic variance: power vs. runtime (log scale); $\sigma=33.6$W validates non-uniform state costs $\mathcal{C}_s$.}
    % \vspace{-6mm}

    \label{fig:combined_results}

\end{figure}

\subsubsection*{\textbf{Findings}} 
\emph{Validating $\mathcal{P}$ (Path Dependency)}
\label{par:path_dependency}
Fig.~\ref{fig:combined_results}(a) shows energy Coefficient of Variation ($CV = \sigma/\mu$) across 39,744 solutions. The right-skewed distribution reveals a deterministic peak ($CV \approx 0$) and heavy tail: 15.6\% exhibit $CV > 0.1$, with extreme cases varying $10\times$ for identical code validating the need for stochastic analysis (Eq.~\ref{eq:expected_energy}); single ``average energy'' values mask order-of-magnitude variations driven by probabilistic path selection $\mathcal{P}$.

To confirm variance stems from path selection rather than data scaling, we analyzed energy variance ($\sigma_E$) vs. instruction variance ($\sigma_{\text{inst}}$). Table~\ref{tab:instruction_variance} shows high-CV solutions exhibit massive instruction variance (e.g., $\sigma_{\text{inst}} \approx 5.5 \times 10^6$), confirming inputs drive execution along different paths $\pi$. For Problem \texttt{p00189}, sparse graphs activate 50\% of vertices while dense graphs force full traversal, explaining the 1.93$\times$ energy variance.

\noindent \emph{Validating Structural Topology (Optimization Potential)}
Fig.~\ref{fig:combined_results}(b) shows energy reduction potential across problems. Over 40\% offer $>90\%$ reduction through topology selection. For Problem \texttt{p02907}, the worst solution (\qty{21.1}{\joule}) vs. best (\qty{0.03}{\joule}) shows \textbf{705$\times$} reduction. This validates deterministic optimization (Sec.~\ref{eq:bounds}): identifying minimum-energy graph structures yields orders-of-magnitude gains.

\begin{table}[h]
    \centering
    % \vspace{-2mm}
    \caption{Correlation of Energy and Inst. Variance (Top 5)}
    % \vspace{-4mm}
    \label{tab:instruction_variance}
    \begin{adjustbox}{width=0.8\columnwidth}
    \begin{tabular}{l|c|r|r}
    \toprule
    \textbf{Solution ID} & \textbf{Energy CV} & \textbf{$\sigma_E$ (J)} & \textbf{$\sigma_{\text{inst}}$} \\
    \midrule
    p03607\_07bc... & 10.10 & 0.0037 & 20,716 \\
    p02959\_1ef7... & 10.10 & 0.0025 & 77,836 \\
    p02614\_6cae... & 10.06 & 0.2583 & 5,520,196 \\
    p02357\_8b1d... & 10.00 & 0.0039 & 23,885 \\
    p00670\_18d2... & 10.00 & 0.0028 & 21,736 \\
    \bottomrule
    \end{tabular}
    \end{adjustbox}
    % \vspace{-2mm}
\end{table}

% \vspace{-2mm}
\noindent \emph{Thermodynamic Validation: Non-Uniform State Costs}
\label{subsec:thermodynamic}
A common simplification in green software engineering assumes Energy $\propto$ Time, implying constant power consumption across all computational operations. The \efg model explicitly rejects this by assigning distinct state costs $\mathcal{C}_s(v)$ based on thermodynamic properties of operations (\eg{} vector instructions vs. memory stalls). This design choice validates the formal separation of state costs ($\mathcal{C}_s$) and transition costs ($\mathcal{C}_t$) in Definition~\ref{def:efg}.
Fig.~\ref{fig:combined_results}(c) shows power vs. runtime. Vertical dispersion demonstrates significant power variance ($\sigma = 33.6$W, 17\% CV) even at fixed runtimes, proving distinct states dissipate energy at different rates. This validates non-uniform state costs $\mathcal{C}_s$ over time-based profiling.

% \vspace{-2mm}
\noindent \emph{Theoretical Implications}
\label{subsec:implications}
The empirical evidence validates \efg tuple components (Section~\ref{def:efg}): (1) High CV values prove transition probabilities $\mathcal{P}$ dominate for 15.6\% of algorithms, validating stochastic analysis. (2) 17\% power variance confirms states have distinct thermodynamic costs independent of time. (3) 705$\times$ structural gaps validate topological optimization (Sec.~\ref{eq:bounds}) as critical.

\subsection{Case Study 2: AI Pipeline Optimization}
\label{subsec:rq2}

\textbf{Application Context.} AI practitioners face a combinatorial explosion when selecting optimization combinations across pipeline stages. With 22 available optimizations yielding $2^{22} \approx 4.2$ million possible combinations, exhaustive testing is prohibitive. \efg addresses this by modeling pipelines as hierarchical state-transition systems where the multiplicative cascade model (Eq.~\ref{eq:multiplicative_model}) predicts combined savings from just 22 individual measurements, eliminating the need for exhaustive testing.

\noindent \textbf{Experimental Setup.} We use data from Rajput et al.~\cite{rajput2025tu} fine-tuning ModernBERT-base~\cite{warner2024smarter} on BigVul~\cite{fan2020ac}, evaluating 22 single-knob variants and 8 multi-knob combinations monitored via CodeCarbon~\cite{benoitcourty2024}. We model the pipeline as a macro-scale \efg with five states (\textit{Data}, \textit{Model}, \textit{Training}, \textit{System}, \textit{Inference}), each with state costs $\mathcal{C}_s$ and transition costs $\mathcal{C}_t$.

\noindent \textbf{Validation.} We apply Eq.~\ref{eq:multiplicative_model} to predict savings for four multi-knob variants (V1--V4) combining optimizations across stages. Predicted values are obtained by multiplying individual savings measured in isolation; observed values are empirical measurements. This tests whether \efg forecasts energy without exhaustive testing (Table~\ref{table:validation}).

\noindent \textbf{Findings.} Table~\ref{table:validation} confirms \efg's predictive utility. V1 shows near-perfect agreement (-0.6\%), confirming true orthogonality between inference engine, gradient checkpointing, LoRA, and FP16. V2 (+1.2\%) and V3 (+2.4\%) show strong alignment with minor noise, indicating slight resource interaction. V4's larger delta (+5.1\%) reveals non-linear hardware effects from compilation and precision reduction, likely thermal throttling or bandwidth saturation. These deviations are diagnostic: practitioners can identify optimization combinations requiring resource allocation adjustments. The near-zero delta in V1 confirms true orthogonality, providing a benchmark for ideal combinations.

\begin{table}[h]
\centering
% \vspace{-3mm}
\caption{Multiplicative Cascade Model Validation}
% \vspace{-3mm}
\label{table:validation}
\begin{adjustbox}{width=1\columnwidth}
\begin{tabular}{l|c|c|c}
\textbf{Variant (Combined strategies)} & \textbf{Predicted} & \textbf{Observed} & \textbf{$\Delta$} \\ \toprule
V1 (InfEngine+GradCkpt+LoRA+FP16) & 86.0\% & 86.6\% & -0.6\% \\ 
V2 (Prune+Compile+FP16) & 74.9\% & 73.7\% & +1.2\% \\ 
V3 (Prune+SeqLen+Compile) & 83.2\% & 80.8\% & +2.4\% \\ 
V4 (Compile+FP16) & 51.1\% & 46.0\% & +5.1\% \\ \bottomrule
\end{tabular}
\end{adjustbox}
% \vspace{-4mm}
\end{table}

% \vspace{-2mm}
\section{Threats to Validity}
\textbf{Internal Validity:} Simulation abstracts thermal effects and OS noise. However, this is deliberate to isolate algorithmic energy costs from environmental noise. \textbf{External Validity:} Results are specific to the simulated Zen 4 architecture, though path-dependency is an architecture-agnostic property of control flow. \textbf{Construct Validity:} The dataset focuses on algorithmic kernels; production software may exhibit different I/O characteristics, a target for future work.

% \vspace{-2mm}
\section{Related Work}

Our work synthesizes formal behavioral modeling, worst-case analysis, and thermodynamic computing across three research streams.

\textbf{Energy-Annotated Behavioral Models.} 
Duarte~\etal{} use Labeled Transition Systems (LTS) to analyze software~\cite{duarte2019model}, annotating transitions with action costs(\eg{} method calls).
Baier~\etal{} apply Probabilistic Model Checking to MDPs with state--transition rewards to verify energy-utility trade-offs~\cite{baier2014probabilistic}.
These approaches exhibit critical limitations that \efg addresses. First, their \textit{additive cost models} ($\sum \text{cost}_i$) treat energy as linearly accumulated, lacking mechanisms to predict optimization interactions. \efg introduces the \textit{Multiplicative Cascade Model} to forecast combined effects of orthogonal optimizations without exhaustive testing. 
Second, 
prior work associates costs with either functional actions~\cite{duarte2019model} or abstract rewards~\cite{baier2014probabilistic} without physical grounding. \efg explicitly separates \textit{State Cost} ($\mathcal{C}_s$, computational residence) from \textit{Transition Cost} ($\mathcal{C}_t$, reconfiguration penalty)---critical for modern hardware where data movement and context switches often dominate computational costs. 
For instance, cache misses and branch mispredictions can incur substantial transition penalties beyond statement execution, which existing models fail to capture. 
Third, prior work targets specific domains (software traces~\cite{duarte2019model}, verification constraints~\cite{baier2014probabilistic}) and suffers state-space explosion for large systems~\cite{baier2014probabilistic}. \efg addresses both limitations through domain-agnostic formulation and \textit{Hierarchical Composition} (Eq.~\ref{eq:hierarchical}), which abstracts sub-graphs as macro-states to enable scalable reasoning from algorithmic kernels to AI pipelines while preserving energy semantics.

\textbf{Worst-Case and Hardware-Aware Analysis.} Worst-Case Energy Consumption (WCEC) extends WCET using CFGs and ILP for safe upper bounds~\cite{jayaseelan2006estimating}. W{\"a}gemann~\etal{}~\cite{wagemann2018whole} advanced this with Power-State-Transition Graphs (PSTG), for hardware peripheral states.
\efg generalizes from hardware device states to algorithmic states. 
Beyond WCEC's deterministic bounds, \efg supports \textit{dual-mode analysis}: deterministic for bounds (Eq.~\ref{eq:path_energy}) and stochastic for expected values (through Bellman equations~\cite{bellman1952theory,bellman1966dynamic}), bridging safety-critical and average-case optimization.

\textbf{Data-Driven Profiling \& Thermodynamic Theory.} Tools like AMALTHEA~\cite{gomes2022model} and CodeCarbon~\cite{codecarbon} employ trace-based regression for energy estimation. These ``black-box'' methods measure \textit{what} happened but cannot structurally explain \textit{why}, and fail to \textit{extrapolate} to unseen execution paths. 
\efg provides a ``white-box'' structural framework for optimization.
Landauer's principle~\cite{landauer1961irreversibility} and reversible computing establish physical energy bounds. \efg operationalizes these by modeling transition costs as thermodynamic expenses, 
combining empirical measurement with theoretical foundations for predictive optimization.

% \vspace{-2mm}
\section{Implications}

\efg offers key implications across the computing ecosystem. For \textbf{software engineers}, it bridges traditional program representations with energy semantics, enabling developers to reason about consumption using familiar graph-based abstractions while incorporating thermodynamic principles. This enables energy-aware programming beyond post-hoc profiling.
For \textbf{AI practitioners}, the predictive optimization methodology addresses combinatorial explosion by leveraging multiplicative modeling---particularly valuable when optimization spaces grow exponentially. Forecasting optimization interactions without exhaustive testing enables strategic energy planning in resource-constrained environments.
For \textbf{researchers and tool developers}, \efg establishes energy complexity alongside time and space complexity, opening research directions in energy-bounded computation program analysis. Its compositional nature supports modular analysis across system boundaries, from programs to ML pipeline optimization.
The framework's cross-domain applicability suggests broader utility for system architects building energy-transparent systems, enabling consistent reasoning from micro-scale operations to macro-scale workflows.

% \vspace{-2mm}
\section{Future Plans}

To advance \efg from its current foundation to a comprehensive methodology, we plan to pursue three directions. First, \textbf{toolchain development} will create production-ready tooling including automated \efg construction from source code, IDE integration for real-time energy feedback, and compiler plugins for energy-aware code generation, making \efg practical for everyday development workflows. 
Second, \textbf{theoretical extensions} will develop formal methods to verify energy equivalence between program versions and composition rules, maintaining energy guarantees when building systems from components. 
Third, \textbf{empirical validation} will conduct large-scale studies across diverse hardware platforms and application domains, using hardware simulators~\cite{binkert2011gem5,heirman2012sniper} for deterministic cost model calibration and extending validation to distributed systems and edge computing. These efforts will transform \efg from a conceptual framework into an ecosystem for energy-aware computing readily adoptable across computational domains.

% \vspace{-2mm}
\section{Conclusions}

This paper introduces \textbf{Energy Flow Graph (\efg)}, a novel computational model bridging graph theory and thermodynamics with software engineering to analyze and optimize energy consumption. \efg captures the path-dependent nature of energy use through state and transition costs, enabling static analysis of optimal paths and predictive modeling of optimization combinations. Our case studies demonstrate \efg's utility from software optimization to AI pipeline management, showing how energy savings propagate across system boundaries. The model establishes energy complexity as a fundamental consideration, enabling both theoretical foundations and practical methodologies for systematic energy optimization across the software, including AI-enabled software, development and deployment lifecycle.

\bibliographystyle{ACM-Reference-Format}
\bibliography{ref}
% \balance

\end{document}